\documentclass[12pt,preprint]{aastex}

\shorttitle{Galaxy Clusters with Thermal Conduction}
\shortauthors{Zakamska \& Narayan}

\def\lsim{\mathrel{\mathpalette\@versim<}}
\def\gsim{\mathrel{\mathpalette\@versim>}}

\begin{document}

\title{Models of Galaxy Clusters with Thermal Conduction}

\author{Nadia L. Zakamska\altaffilmark{1}}
\email{nadia@astro.princeton.edu}
\author{Ramesh Narayan\altaffilmark{1,2}}
\email{rnarayan@cfa.harvard.edu}
\altaffiltext{1}{Department of Astrophysical Sciences, Princeton University, Princeton, NJ 08544}
\altaffiltext{2}{Harvard-Smithsonian Center for Astrophysics, 60 Garden Street, Cambridge, MA 02138}

\begin{abstract}
We present a simple model of hot gas in galaxy clusters, assuming
hydrostatic equilibrium and energy balance between radiative cooling
and thermal conduction.  For five clusters, A1795, A1835, A2199, A2390
and RXJ1347.5-1145, the model gives a good description of the observed
radial profiles of electron density and temperature, provided we take
the thermal conductivity $\kappa$ to be about 30\% of the Spitzer
conductivity.  Since the required $\kappa$ is consistent with the
recent theoretical estimate of Narayan \& Medvedev (2001) for a
turbulent magnetized plasma, we consider a conduction-based
equilibrium model to be viable for these clusters.  We further show
that the hot gas is thermally stable because of the presence of
conduction.  For five other clusters, A2052, A2597, Hydra A, Ser
159-03 and 3C295, the model requires unphysically large values of
$\kappa$ to fit the data.  These clusters must have some additional
source of heat, possibly an active galactic nucleus since all the
clusters have strong radio galaxies at their centers.  We suggest that
thermal conduction, though not dominant in these clusters, may
nevertheless play a significant role by preventing the gas from
becoming thermally unstable.
\end{abstract}


\section{Introduction}

For many years it was thought that radiative losses via X-ray emission
in galaxy clusters leads to a substantial inflow of gas, and mass
dropout, in the form of a ``cooling flow'' (see \citealt{fabi94} for a
review).  Mass deposition rates were estimated to be as much as
several hundred $M_\odot{\rm yr^{-1}}$ in some clusters (e.g.,
\citealt{pere98}).  However, there was little direct evidence for the
mass dropout in any band other than X-rays \citep{fabi94}.

Recent high resolution X-ray data from {\it XMM-Newton} and {\it
Chandra} indicate that there is no evidence for the multi-temperature
gas that one expects if there is substantial mass dropout
(\citealt{pete01, tamu01, bohr01, fabi01, mole01, mats02, john02}).
The new data strongly suggest that mass dropout must be prevented by
some additional source of heat that balances radiative losses.  Two
possibilities are currently being investigated for the heat source:
(i) energy in jets, outflows, and radiation from a central active
galactic nucleus (\citealt{pedl90, tabo93, chur00, chur02, ciot01,
brue02}), and (ii) thermal energy from outer regions of the cluster
transported to the central cooling gas by conduction
(\citealt{nara01}, hereafter NM01).

Conduction in clusters has been discussed by several authors over the
years in various contexts \citep{binn81, tuck83, bert86, breg88,
gaet89, rosn89, davi92, pist96}, but its importance was always
considered doubtful.  For conduction to have any significant effect on
the cooling gas in a cluster, the effective isotropic coefficient of
conduction $\kappa$ has to be a reasonable fraction of the classical
\citet{spit62} conductivity $\kappa_{Sp}$.  However, conventional
wisdom is that magnetic fields strongly suppress conduction
perpendicular to the field.  Therefore, while conduction may be very
efficient parallel to the field, the overall isotropic $\kappa$ is
expected to be $\ll\kappa_{Sp}$.

This picture has changed with the recent work of NM01 who, following
earlier work by \citet{rech78}, \citet{chan98}, \citet{chan99} and
\citet{maly01}, showed that a turbulent MHD medium in which the
fluctuation spectrum extends over two or more decades of spatial
scales has an effective $\kappa$ that is a substantial fraction of
$\kappa_{Sp}$.  This is because of chaotic transverse wandering of
field lines as a result of strong MHD turbulence \citep{gold95}, which
leads to a large enhancement of cross-field diffusion.  NM01 estimated
that the ratio $f=\kappa/\kappa_{Sp}$ in a turbulent MHD medium is
approximately $0.2$, though, given the uncertainties in their model,
the value could probably lie anywhere in the range $\sim0.1-0.4$.

NM01 showed via a simple order-of-magnitude estimate that thermal
conduction with $f\sim0.2$ is sufficient to balance the radiative
losses of the cooling gas in X-ray clusters.  Based on this, they
suggested that conduction might be a large part of the explanation for
the lack of mass dropout in X-ray clusters.  Their suggestion finds
support in the work of \citet{voig02} and \citet{fabi02} who show that
conduction with $f$ in the range $0.1-1$ can indeed be a dominant heat
source in X-ray clusters except perhaps in the very central regions.

The aim of the present paper is to study conduction in galaxy clusters
in more detail.  Our goals are two-fold.  First, we consider a very
simple model in which the gas is in hydrostatic equilibrium and in
thermal balance, with cooling exactly compensated by heat conduction.
We investigate whether this basic model can fit the observed shapes of
the electron density profile $n_e(r)$ and temperature profile $T(r)$
of clusters for which high-resolution observations are available.  We
believe this approach, which is complementary to that of
\citet{voig02}, is a good test of the conduction hypothesis, and a
reliable method of estimating what value of $f$ is needed in various
clusters to explain the observations.  Second, having found a steady
state model for a given cluster, we check whether the hot gas is
thermally stable. We briefly discuss the importance and plausibility
of other energy sources.

The paper is organized as follows.  In \S2 we write down the relevant
differential equations and boundary conditions for a steady
equilibrium model of hot gas in a cluster.  In \S3 we compare model
predictions with data on 10 clusters, and in \S4 we discuss the
thermal stability of the gas in these clusters.  We conclude with a
discussion in \S5.  Throughout the paper, we take $H_0=70$ km
sec$^{-1}$ Mpc$^{-1}$, $\Omega_M=0.3$ and $\Omega_{\Lambda}=0.7$,
rescaling observational results whenever the original papers have used
a different cosmology for their analysis.

\section{The Model}

We consider a spherically symmetric cluster containing hot gas in
equilibrium with no inflow.  The condition of hydrostatic equilibrium
gives
\begin{equation}
{1\over\rho}{dp\over dr}=-{d\phi\over dr}, \label{hydro}
\end{equation}
where $\rho(r)$ is the gas density, $p(r)$ is the pressure, and
$\phi(r)$ is the gravitational potential.  We assume that gas pressure
dominates and that magnetic fields have a negligible dynamical effect,
so that
\begin{equation} 
p = {\rho k_BT\over\mu m_u} = {\mu_e\over\mu}n_ek_BT
\equiv \rho c_s^2, \label{pressure}
\end{equation}
where $T(r)$ is the temperature of the gas, $n_e(r)$ is the electron
number density, $\mu$ is the mean molecular weight, $\mu_e$ is the
molecular weight per electron, $m_u$ is the atomic mass unit, and
$c_s(r)$ is the isothermal sound speed of the gas.  We assume that the
gas has hydrogen fraction $X=0.7$ and helium fraction $Y=0.28$, so
that $\mu=0.62$, $\mu_e=1.18$ and $\mu_e/\mu=1.91$ (taking hydrogen
and helium to be fully ionized).

The energy equation of the gas is a simple balance between radiative
cooling and heating by thermal conduction.  Thus, in steady state,
we have
\begin{equation}
{1\over r^2}{d\over dr}(r^2F) = -j, \label{energy}
\end{equation}
where $F(r)$ is the heat flux due to electron conduction and $j(r)$ is
the rate of radiative cooling per unit volume of the gas.  We assume
that the conductivity of the gas $\kappa$ is a fixed fraction $f$ of
the classical \citet{spit62}  conductivity.  Thus,
\begin{equation}
F=-\kappa{dT\over dr}, \label{heatflux}
\end{equation}
\begin{equation}
\kappa = f\kappa_{Sp} = f {1.84\times10^{-5}T^{5/2}\over\ln\Lambda}
~{\rm erg\,s^{-1}\,K^{-1}\,cm^{-1}}, \label{Spitzer}
\end{equation}
where $\ln\Lambda\sim37$ is the usual Coulomb logarithm.  In the
relevant range of temperatures the gas cools primarily by thermal
bremsstrahlung.  We therefore write \citep{rybi79}
\begin{equation}
j = 2.1\times10^{-27}n_e^2T^{1/2} ~{\rm erg\,cm^{-3}\,s^{-1}}, \label{cooling}
\end{equation}
where the numerical coefficient corresponds to the particular gas composition
that we have assumed; the coefficient also includes an average Gaunt
factor.

The gravitational acceleration $d\phi(r)/dr$ due to the mass in the
cluster is obtained by solving Poisson's equation,
\begin{equation}
{1\over r^2}{d\over dr}\left(r^2{d\phi\over dr}\right) = 
4\pi G(\rho_{DM}+\rho), \label{Poisson}
\end{equation}
where $\rho_{DM}(r)$ is the density distribution of the dark matter,
which we assume has the NFW form \citep{nava97} with a softened core:
\begin{equation}
\rho_{DM}(r)={M_0/2\pi \over (r+r_c)(r+r_s)^2}. \label{NFW}
\end{equation}
Here, $r_s$ is the standard scale radius of the NFW profile, $r_c$ is
a core radius inside which the density becomes constant, and $M_0$ is
a characteristic mass (e.g., if $r_c=0$, the mass inside a sphere of
radius $r_s$ is $(2\ln2-1)M_0$).  

We relate the parameters of the NFW profile to the temperature
$T_{out}$ of the hydrostatic gas at $r\sim r_s$, well outside the
cooling region of the cluster. We use the mass-temperature relation
from \citet{afsh02} and the mass-scale relation from \citet{maoz97},
so that, once we know $T_{out}$, we can estimate the virial mass
$M_{vir}$ and $r_s$.  The former is defined as the mass inside the
radius within which the mean density of the dark matter is
$200\rho_{crit}(z)$, where $\rho_{crit}(z)$ is the critical density of
the universe at the redshift of the cluster: $\rho_{crit}(z)=3
H(z)^2/8\pi G$.  The normalization of the NFW profile is then obtained
via a simple calculation:
\begin{equation}
M_0=2\pi \rho_{crit}(z) r_s^3 \left( \frac{200}{3}\right)
\frac{c^3}{\ln(1+c)-c/(1+c)},
\end{equation}
where $c$ is the concentration parameter:
\begin{equation}
c=\frac{1}{r_s}\left(\frac{3 M_{vir}}{4 \pi \cdot 200 \rho_{crit}(z)}\right)^{1/3}.
\end{equation}
It is interesting that simulations of cluster formation allow us
almost completely to determine the cluster potential from a single
observed number, $T_{out}$.  The only remaining uncertainty is the
shape of the density distribution in the very inner regions of the
cluster, which we parametrize through the core radius $r_c$.

The value of the core radius $r_c$ is somewhat uncertain.  It is zero
in the original NFW model, but cluster lensing studies \citep{tyso98,
shap00} suggest that it may be tens of kpc in some clusters.  As
explained in \S3 below, we try two values of $r_c$: $r_c=0$ and
$r_c=r_s/20$.

Equations (\ref{hydro}), (\ref{energy}), (\ref{heatflux}) and
(\ref{Poisson}) are four ordinary differential equations for the four
variables $n_e(r)$, $T(r)$, $F(r)$ and $d\phi(r)/dr$.  We solve these
equations numerically between $r=0$ and $r=2r_s$ with the following
four boundary conditions.  Two boundary conditions are taken directly
from the observations of a given cluster, namely the temperature
$T_{in}=T(0)$ at the cluster center and $T_{out}=T(2r_s)$.  For
$T_{out}$ we use the observed temperature at around $r\sim r_s$, which
is well outside the cooling region in the center.  The other two
boundary conditions are $F(0)=0$ and $M(0)= 0$, where $M(r)$ is the
the mass (dark matter plus gas) inside radius $r$. 

We note that we make a slight approximation by assuming a perfect steady
state.  In real clusters with conductive heating, the energy for the
cooling ultimately comes from gravity and so the gas in the cluster
must slowly settle in the gravitational potential.  This effect would
be accentuated by conduction to the outside, as discussed by
\citet{loeb02}.  We do not include gravitational settling in our
model.  Instead, we assume that the outer regions of the cluster
behave like an infinite isothermal heat source at a temperature of
$T_{out}$.  We have experimented with different radii for the outer
boundary of our models, from $r=r_s$ to $r=5r_s$, and find that the
results are not sensitive to this choice.  The models discussed below
correspond to $r=2r_s$.

\section{Results}

We have analysed the following 10 clusters for which high angular resolution
data are available from {\it Chandra} or {\it XMM-Newton} or both:
A1795, A1835, A2052, A2199, A2390, A2597, Hydra A, RXJ 1347.5-1145,
Ser 159-03, and 3C295.  Table 1 lists the relevant parameters for each
cluster.  In the case of several of these clusters, deprojected
density and temperature profiles are available, and we have used them
in our modeling.  In a few cases, however (A2597, A2052, Hydra A, Ser
159-03), the original authors have not published deprojected profiles,
and we had to use the un-deprojected data.  Given the observational
uncertainties, deprojection only affects the profiles in the innermost
regions of the cluster, within about 30 kpc. On the other hand, our
model solutions only depend slightly on the inner temperature.
Therefore, we believe that the shapes of the profiles are essentially
the same outside about 30 kpc, whether we use deprojected or
un-deprojected temperatures.

Table 1 lists for each cluster the inner and outer temperatures,
$T_{in}$ and $T_{out}$, which were used as boundary conditions in the
modeling. It also lists the parameters of the NFW profile of the dark
matter, which were estimated from $T_{out}$ as described in \S2.  The
last column gives references to the original publications for the
X-ray data.

We carried out the modeling as follows.  We first assumed that the
core radius $r_c=0$ and selected a value of $f$ (say 0.2).  We
adjusted the inner density $n_{in}=n_e(0)$ of the gas such that on
integrating the four differential equations (\ref{hydro}),
(\ref{energy}), (\ref{heatflux}) and (\ref{Poisson}) from $r=0$ to
$r=2r_s$, the model temperature agreed with the measured outer
temperature $T_{out}$.  We then compared the model profiles of
$n_e(r)$ and $T(r)$ with the data; we restricted ourselves to a
comparison by eye since neither the deprojected data nor the models
are reliable enough for a detailed $\chi^2$ fit.  If the fit was not
satisfactory, we tried other values of $f$.  If no value of $f$ gave a
satisfactory fit, we repeated the process by setting $r_c=r_s/20. $
(The ratio 20 was selected to be consistent with estimates from
lensing; the solutions are insensitive to the exact value.)  For most
clusters, we obtained a good fit for some choice of $f$ and one of the
two values of $r_c$ we tried. The values of the best-fit parameters
are listed in Table 2.

We consider a model to be acceptable if the model profiles of $n_e(r)$
and $T(r)$ agree well with the data and if the required value of $f$
lies within the range $f\sim0.1-0.4$ predicted by theory (NM01).  By
this criterion, 5 of the 10 clusters we studied are consistent with a
pure thermal conduction model: A1795, A1835, A2199, A2390, and RXJ
1347.5-1145.  Figures 1, 2 and 3 show the fits for three of these
clusters (A1795, A2199 and A2390).

We consider the good agreement between the model and the data
encouraging since the model has very few adjustable parameters.  Note
in particular that our primary adjustable parameter $f$ affects only
the overall normalization of $n_e(r)$, but has no effect on the shapes
of either $n_e(r)$ or $T(r)$.  The fact that the model curves agree
well with the data is thus a significant result.  To our knowledge,
this kind of modeling of the shapes of profiles has not been reported
before.  There is a slight inconsistency in the density profile below
10 kpc in A2199 (and also RXJ 1347.5-1145), but this is hardly
surprising since the central galaxy is likely to introduce large
perturbations to our assumed NFW profile in this region.

Of the 5 clusters for which we obtain acceptable models, we note that
A1835 is a complicated case because the {\it Chandra} and {\it
XMM-Newton} data do not agree with each other; indeed, \citet{mark02}
has shown that there are problems with the analysis of both datasets.
We obtain a reasonable model for this cluster if we use the density
profile from {\it XMM-Newton} data \citep{maje02} corrected for
flares.  For A1795, although both {\it Chandra} and {\it XMM-Newton}
data sets are available \citep{etto02, tamu01} we do not use the {\it
XMM-Newton} temperature profile because it is un-deprojected.

The remaining 5 clusters, A2052, A2597, Hydra A, Ser 159-03 and 3C295,
all require large values of $f$, generally greater than unity.  Since
such values are unphysical, we conclude that these clusters must have
some other source of energy in addition to thermal conduction.
Heating from a central active galactic nucleus is a promising
possibility \citep{chur02, brue02}.  The idea is supported by the fact
that all these clusters possess strong active galactic nuclei (AGN) in
their centers which are also all powerful extended radio sources
(3C317 in A2052, PKS2322-122 in A2597, 3C218 in Hydra A, PMNJ2313-4243
in Sersic 159-03, and the cluster 3C295 is named after the radio
source in its center). Strong extended radio emission is indicative of
powerful outflows interacting with the intracluster medium
(e.g. \citealt{brid84}).  In contrast, of the 5 well-fit clusters,
only two (A1795 and A2199) are known to have strong extended radio
sources in their centers (4C26.42 and 3C338, respectively).

Mass inflow and dropout is, of course, another possible source of
energy, as has been discussed for many years in the context of cooling
flows.  However, as mentioned in \S1, spectroscopic analyses of {\it
Chandra} and {\it XMM-Newton} data of several clusters have failed to
find any evidence for the multi-phase gas expected if there is
significant mass inflow and dropout.  The observations place an upper
limit on the mass accretion rate in a few clusters of about 10-20\% of
the nominal value needed to explain the observed X-ray flux
(e.g. \citealt{davi01, schm01, john02}).  Accretion at this level
cannot substantially modify the energy balance, so one needs either
conduction or mechanical energy from an AGN to fit the observations.
We have computed models with larger levels of mass accretion, at the
level of hundreds of $M_{\sun}$ yr$^{-1}$, and we do find that
accretion can then explain the X-ray data.  Although the details of
accretion and energy release during the mass dropout are uncertain
(see different versions of energy balance in \citealt{bert86} and
\citealt{sara88}), we find that we can reproduce the observed density
and temperature profiles with reasonable assumptions.

\section{Thermal Stability}

To analyze the stability of the steady-state solutions described in
\S3, we work with the full time-dependent equations.  Equation
(\ref{hydro}) generalizes to the Euler equation,
\begin{equation}
{\partial{\bf v}\over\partial t} + ({\bf v}\cdot{\bf \nabla})
{\bf v} = -{1\over\rho}{\bf \nabla}p - {\bf \nabla}\phi, \label{Euler}
\end{equation}
where ${\bf v}$ is the gas velocity.  We also have the continuity
equation,
\begin{equation}
{\partial\rho\over\partial t}+{\bf \nabla}\cdot(\rho{\bf v}) = 0.
\label{continuity}
\end{equation}
The energy equation (\ref{energy}) now becomes the time-dependent
entropy equation,
\begin{equation}
\rho T\left[{\partial s\over\partial t} + ({\bf v}\cdot
{\bf \nabla})s\right] = -j -{\bf \nabla}\cdot{\bf F} + q^+,
\label{entropy}
\end{equation}
where the specific entropy $s$ of the gas is defined in terms of the
specific internal energy $\epsilon$ and specific enthalpy $w$ by
\begin{equation}
Tds = d\epsilon - {p\over\rho^2}d\rho = dw - {dp\over\rho}.
\label{entropydef}
\end{equation}
For an ideal gas,
\begin{equation}
\epsilon={1\over(\gamma-1)}c_s^2, \qquad
w={\gamma\over(\gamma-1)}c_s^2, \label{enthalpy}
\end{equation}
where $\gamma$ is the adiabatic index and $c_s$ is the isothermal
sound speed (Eq.~\ref{pressure}).  In our calculations, we set
$\gamma=5/3$.  

For completeness, we have included in equation (\ref{entropy}) a
heating term $q^+$ in addition to the term due to thermal conduction.
This term may refer, for instance, to heating from a central AGN.  We
assume that the heating rate per unit volume has a power-law
dependence on the density,
\begin{equation}
q^+ = q_0^+ \rho^\alpha. \label{heating}
\end{equation}

The heat flux equation (\ref{heatflux}) and the Poisson equation
(\ref{Poisson}) are still valid, except that we generalize them to
include non-spherical perturbations:
\begin{equation}
{\bf F}=-\kappa{\bf \nabla}T, \label{heatflux2}
\end{equation}
\begin{equation}
\nabla^2\phi=4\pi G\left(\rho_{DM}+\rho\right). \label{Poisson2}
\end{equation}

The above five equations, (\ref{Euler}), (\ref{continuity}),
(\ref{entropy}), (\ref{heatflux2}) and (\ref{Poisson2}), describe the
general time-dependent evolution of the gas.  Note that, while we
include the effect of a turbulent magnetic field on conduction through
the parameter $f<1$ in equation (\ref{Spitzer}), we do not include
magnetic pressure or tension in the Euler equation (see \citealt{loew90} 
for a discussion of these effects).

For the stability analysis, we take a steady state spherical solution
of the equations, e.g., one of the cluster models described in \S3,
and consider local linear perturbations of the solution within the WKB
approximation; that is, we assume that the perturbations are small and
are proportional to $\exp(ikx-i\omega t)$.  \citet{balb88} has shown
that radial perturbations, with wavevector $k$ parallel to the local
direction of gravity, are not well-described by a local analysis;
these perturbations also do not show thermal instability. In the
following we therefore consider only tangential perturbations, with
$k$ and the corresponding $x$ taken to be in a tangential direction
with respect to the radius vector.

We carry out the linear stability analysis in the standard way; see
\citet{fiel65} for a classic treatment of the thermal stability
problem.  In brief, we linearize the equations, substitute
perturbations of the form $\exp(ikx-i\omega t)$ and derive the
dispersion relation for the mode frequency $\omega$.  If any of the
solutions for $\omega$ has a positive imaginary part, the
corresponding mode grows with time and the system is unstable.  If all
modes for all allowed values of $k$ have ${\rm Im}(\omega)\le0$, we
say that the system is stable.

Before presenting the results, we note that there are several physical
effects present in this problem, each with a characteristic frequency
or time scale.  In addition to the acoustic frequency $c_sk$, we have
frequencies associated with gravity (Jeans frequency $\omega_J$),
cooling ($\omega_{cool}$) and heating ($\omega_{heat}$), and a
characteristic time scale associated with conduction ($\tau_{cond}$):
\begin{equation}
\omega_J=(4\pi G\rho)^{1/2}, \quad
\omega_{cool}={j\over\rho c_s^2}, \quad
\omega_{heat}={q^+\over\rho c_s^2}, \quad
\tau_{cond}={\kappa T\over\rho c_s^4} .
\end{equation}
Each of the parameters, $\omega_J$, $\omega_{cool}$, $\omega_{heat}$
and $\tau_{cond}$, has been defined in such a way that the parameter
goes to 0 as the corresponding physical effect becomes negligible.

We discuss first a simple case in which we assume that conduction is
negligible ($\tau_{cond}\to 0$) and, for simplicity, we ignore
self-gravity ($\omega_J \to 0$).  Energy balance of the equilibrium
system requires $q^+=j$, and thus $\omega_{heat}=\omega_{cool}$.
Assuming that $q^+$ and $j$ scale with density and temperature as in
equations (\ref{heating}) and (\ref{cooling}), and taking
$\gamma=5/3$, we then obtain the following dispersion relation for
$\omega$:
\begin{equation}
(-i\omega)^3+{1\over3}\omega_{cool}(-i\omega)^2+{5\over3}c_s^2k^2(-i\omega)
=\left(1-{2\over3}\alpha\right)\omega_{cool}c_s^2k^2.
\label{dispersion1}
\end{equation}
By inspection we see that one of the solutions for $\omega$ is always
positive imaginary if $\alpha<1.5$.  Thus, if $\alpha<1.5$, the system
is always thermally unstable.  Physically, we expect many heating
mechanisms to be described by $\alpha$ in the range between 0
(constant heating rate per unit volume) and 1 (constant heating rate
per unit mass).  According to equation (\ref{dispersion1}), such
systems will be thermally unstable.  This suggests that models which
try to solve the cooling flow problem by invoking local heating
\citep{chur02, brue02} may still have some difficulties.  While they
allow equilibrium configurations, the equilibria they obtain are
likely to be unstable.  Unless the heating occurs preferentially in
high density regions, with an index $\alpha\ge1.5$ (which seems
implausible), these models require thermal conduction or some other
agency for stability.

We move next to the problem of interest to this paper, namely models
in which cooling is balanced entirely by heat conduction ($q^+=0,
~\omega_{heat}=0$).  The dispersion relation now takes the form
(again setting $\gamma=5/3$)
\begin{eqnarray}
&{3\over2}\omega^3+\omega^2\left(\frac{1}{2}i \omega_{cool} + i
k^2 c_s^2 \tau_{cond}\right) +\omega \left(- {5\over2} k^2
c_s^2+{3\over2}\omega_J^2\right)+\nonumber\\
&+\frac{3}{2}i\omega_{cool}k^2 c_s^2+\frac{1}{2}i
\omega_{cool}\omega_J^2-ik^4 c_s^4\tau_{cond}+i
k^2 c_s^2\tau_{cond}\omega_J^2 =0 \label{dispersion2}
\end{eqnarray}
Analysis of this dispersion relation shows that an instability is present
only when the condition
\begin{equation}
k^2 c_s^2 <
\frac{1}{2}\left(\omega_J^2+\frac{3}{2}\frac{\omega_{cool}}{\tau_{cond}}+
\sqrt{\omega_J^4+5\frac{\omega_J^2\omega_{cool}}{\tau_{cond}}+\frac{9}{4}
\frac{\omega_{cool}^2}{\tau_{cond}^2}}\right)\label{instability}
\end{equation}
is satisfied.

We see that, as expected (e.g., Medvedev \& Narayan 2002), conduction
eliminates the thermal instability for short wavelength modes.
Physically, the reason is that the shorter the wavelength the larger
the temperature gradient for a given mode amplitude; since conduction
tries to iron out temperature fluctuations and the conductive flux is
proportional to the temperature gradient, short wavelength modes are
stabilized.  For long wavelengths, however, sound waves can no longer
stabilize the gravitational Jeans instability, and (finite) conduction
cannot stabilize the thermal instability. This is the physical meaning
of the upper limit on $k$ given in equation (\ref{instability}).

Let us denote the critical wavevector for instability by $k_{crit}$
and the corresponding wavelength by $\lambda_{crit}=2\pi/k_{crit}$.
The relevant question is the following: can an unstable mode with
wavelength $\lambda_{crit}$ or larger fit within the system?  Recall
that our modes are assumed to be in the tangential direction and we
carried out the analysis under a local WKB approximation.  For this
analysis to be meaningful, the wavelength of the mode must be
substantially smaller than the circumference $2\pi r$.  A given
cluster model is thus stable if $\lambda_{crit}$ is greater than the
largest wavelength for which the WKB analysis is valid, which is
probably around $\pi r$.

Figure 4 shows the variation of $\lambda_{crit}/\pi r$ with $r$ for
the 5 clusters for which we obtained satisfactory models with
conduction (\S3).  We see that all 5 clusters are thermally stable
according to the criterion described above.  One interesting feature
is that all the clusters show very similar minimal values of 
$\lambda_{crit}/\pi r$ which appear to be fairly close to the threshold of
instability.
Such a situation would be natural if the clusters, to begin with, had
more hot gas than they have now and were thermally unstable.  One
expects gas to have cooled and dropped out early in the life of the
cluster so that the system naturally reached, and has since
maintained, a marginally stable configuration.  A global stability
analysis would need to be carried out to check this idea in more
detail.

\section{Discussion}

NM01 found that thermal conduction with $f\equiv\kappa/\kappa_{Sp}$ of
order a few tenths can explain the gross energetics of hot gas in
galaxy clusters.  They showed that the heat flux transported to the
cooler central regions of the cluster from further out is roughly
consistent with what is needed to replace the energy lost through
radiative cooling.  \citet{gruz02} and \citet{fabi02} confirmed this
conclusion.

In this paper, we have carried out a further test of the conduction
model by investigating the shapes of the density and temperature
profiles, $n_e(r)$ and $T(r)$, in an equilibrium cluster.  We solve
for these profiles self-consistently, as compared to previous authors,
e.g., \citet{bert86, breg88, voig02} and \citet{fabi02}, who either
used the observations directly or employed simple analytical
expressions for the shapes of $n_e(r)$ and/or $T(r)$.

Among the 10 clusters that we have studied, 5 clusters, namely, A1795,
A1835, A2199, A2390 and RXJ1347.5-1145, can be fitted well with a pure
conduction model and with values of $f\sim0.2-0.4$ (Table 2, Figs. 1,
2, 3).  Since the model involves no fitting parameters other than $f$,
which is used primarily to set the normalization of $n_e$, and $r_c$,
which is not really a fitting parameter but rather is given one of two
values (\S3), the good agreement between the model profiles of
$n_e(r)$ and $T(r)$ and the data is very encouraging.  For these five
clusters, the best-fit values of $f$ lie within the range expected on
theoretical grounds (NM01).  We believe that these are strong
arguments in support of the conduction model.  We should note,
however, that the theoretical estimate $f\sim0.1-0.4$ obtained by NM01
is based on a specific (plausible) model of the magnetic field
topology, suggested by the work of Goldreich \& Sridhar (1995, 1997),
but the answer might change substantially if the field topology (which
is poorly understood in clusters) is very different.

The other 5 clusters, namely A2052, A2597, Hydra A, Ser 159-03 and
3C295, require larger values of $f$, of order unity or greater, and we
consider them to be inconsistent with a pure conduction model.  It is
interesting that these five clusters all have active nuclei with
strong radio activity, so that nuclear activity might well supply the
additional heat required by our models.  It is unclear exactly how an
AGN would heat the cluster medium.  Turbulent mixing of hot gas in a
jet with the surrounding cooler gas is a possibility \citep{chur02,
brue02}, though the effect of magnetic fields on this mixing is poorly
understood.  Heating of the gas by cosmic rays \citep{loew91} or hard
X-rays \citep{ciot01} from the AGN are other possibilities.

In \S4 we showed by means of a local linear stability analysis that
the five clusters for which we have obtained good fits with a pure
conduction model are all thermally stable.  This is an interesting
result because, without conduction, cooling gas in clusters is
generally thermally unstable.  We find it reassuring that a single
mechanism, namely conduction, is able both to supply the heat lost via
radiative cooling and to control the thermal instability in these
clusters.  In the other five clusters, which require an additional
source of energy such as a central AGN, the thermal stability of the
gas is not assured.  One way of ensuring stability is to have a heat
source with a strong non-linear dependence on density, e.g.,
$\alpha>1.5$ in equation (\ref{heating}); this appears somewhat
unnatural.  Alternatively, conduction, though not an important energy
source in these clusters, might still play a role in controlling the
instability.

Based on the above results, we suggest that AGN heating and conduction
both are important in clusters.  In some clusters, AGN heating
dominates and conduction plays a secondary, though still important,
role in helping to stabilize the system.  In other clusters, AGN
activity is weak, and conduction supplies the energy for cooling as
well as prevents the gas from becoming unstable.

While our work suggests that conduction is effective over much of the
volume of a cluster, we note that {\it Chandra} observations have
revealed sharp temperature jumps in some clusters (e.g.,
\citealt{mark00}; Vikhlinin, Markevitch \& Murray 2001ab).  These cold
fronts are associated with the interface between the intracluster
medium and an infalling galaxy or group.  The observations clearly
indicate that conduction is strongly suppressed across the fronts.
\citet{vikh01b} have argued that the magnetic field is stretched out
parallel to the cold front by the motion of the infalling galaxy/group
and that this explains the very low conductivity.  \citet{mazz02}
present interesting evidence for a possible Kelvin-Helmholtz
instability in the cold front in A3667, confirming that the fronts are
probably transient features with lifetimes of only a couple of
dynamical times.  The presence of these fronts is thus unlikely to
affect the strong conduction that we have hypothesized over the bulk
of the cluster.  If cold fronts survive for longer than the cooling
time of the gas, then the thermally isolated cool gas would experience
runaway cooling.  Since such runaway cooling is ruled out by the
absence of substantial mass dropout, the lifetimes of cold fronts
cannot be longer than the cooling time.

\bigskip
The authors thank Bill Forman, Christine Jones Forman and Larry David
for useful discussions, and the referee for helpful comments on the
manuscript.  RN thanks the National Science Foundation for support
under grant AST 9820686.

\clearpage

\begin{figure}
\plotone{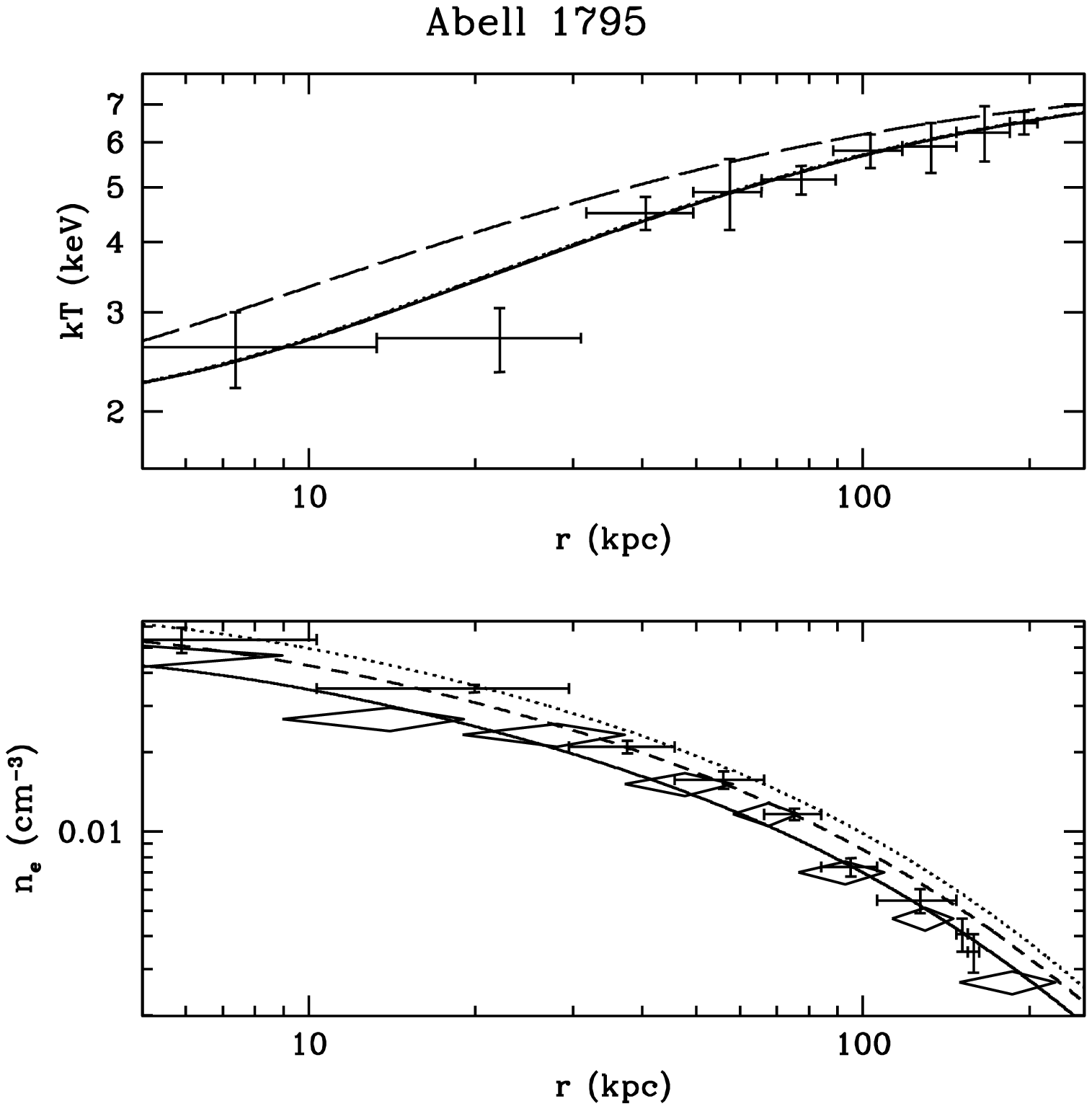}
\figcaption{Observed and modeled temperature and density profiles for
A1795. Crosses correspond to $Chandra$ data (Ettori et al. 2002) and
diamonds to $XMM-Newton$ data (Tamura et al. 2001).  The solid, dashed
and dotted lines refer to models with $r_c=r_s/20$ and different
conductivity coefficients, $f=0.2, ~0.3, ~0.4$, respectively.  Note
that changing the conductivity has no effect on the temperature
profile.  In the case of the density profile, $f$ changes the
normalization but not the shape.  The long-dashed line shows the
temperature profile for a model in which $r_c=0$ and $f=0.4$.}
\end{figure}

\clearpage

\begin{figure}
\plotone{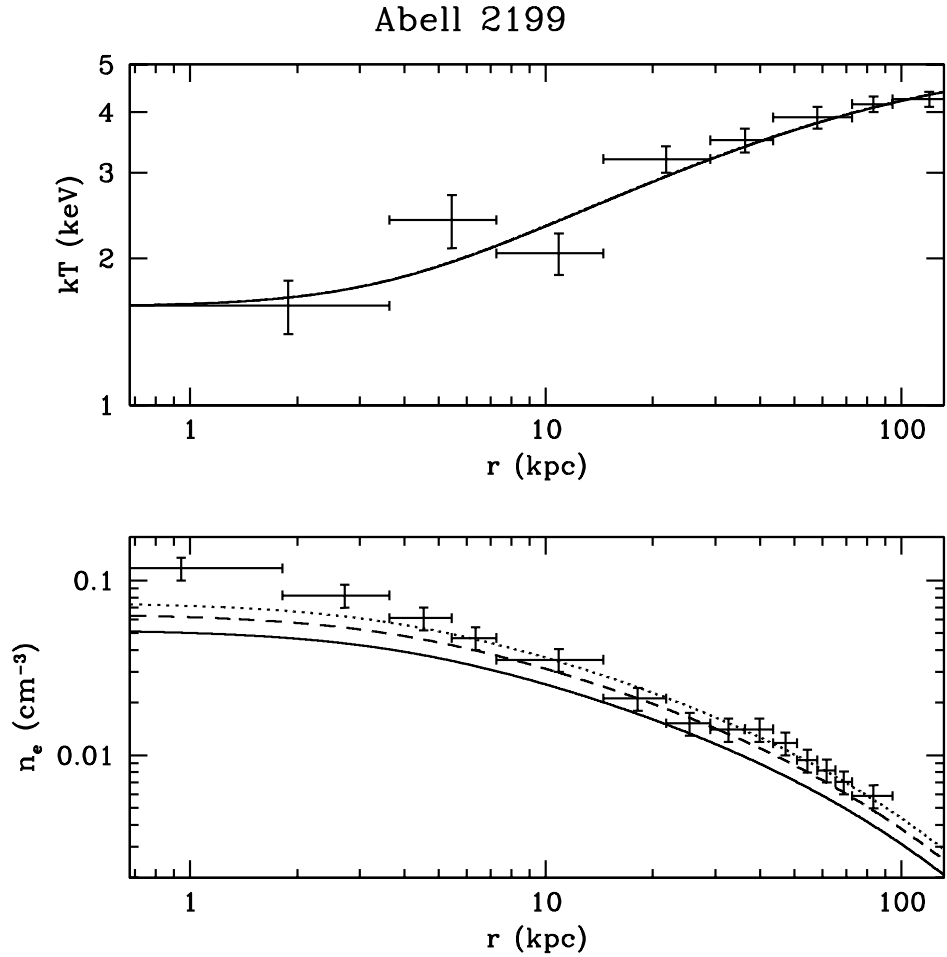}
\figcaption{
Observed and modeled temperature and density profiles for
A2199.  Crosses indicate $Chandra$ data (Johnstone et al. 2002).  The
solid, dashed and dotted lines refer to models with $r_c=0$ and
$f=0.2, ~0.3, ~0.4$, respectively.}
\end{figure}

\clearpage

\begin{figure}
\plotone{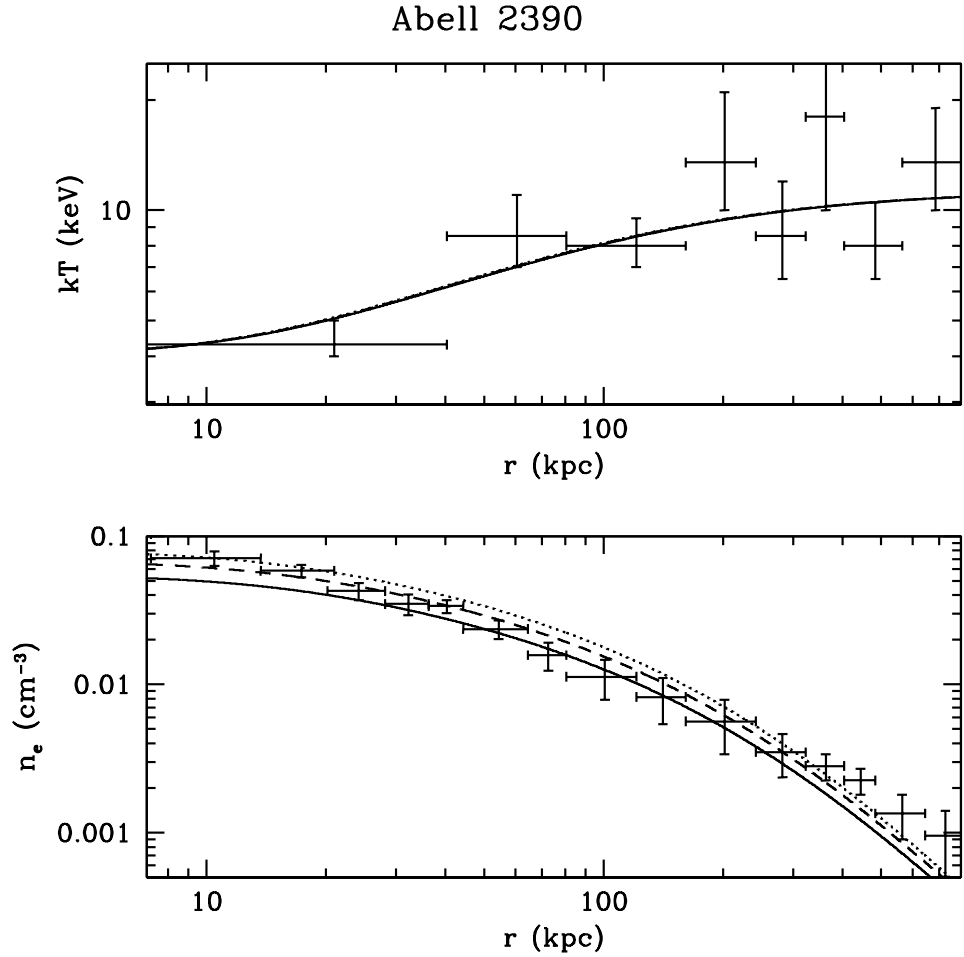}
\figcaption{
Observed and modeled temperature and density profiles for
A2390.  Crosses indicate $Chandra$ data (Allen et al. 2001b).  The
solid, dashed and dotted lines refer to models with $r_c=r_s/20$ and
$f=0.2, ~0.3, ~0.4$, respectively.}
\end{figure}

\clearpage

\begin{figure}
\plotone{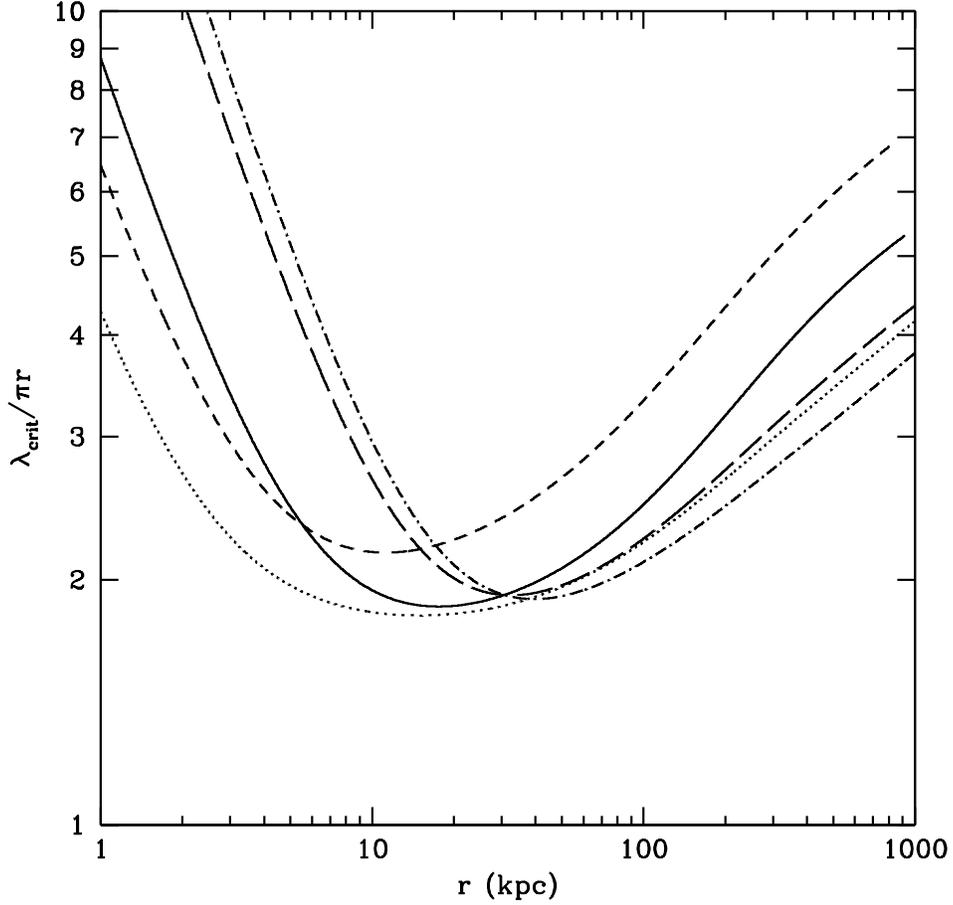}
\figcaption{
Results of a local stability analysis.  The stability
parameter $\lambda_{crit}/\pi r$ is shown as a function of radius for
5 clusters that are fit well by a conduction model.  The solid line
corresponds to A1795, the dotted line to A1835, the short-dashed line
to A2199, the long-dashed line to A2390, and the dotted-dashed line to
RXJ1347.5-1145. Note that all the curves have similar shapes, with
minima around 10-40 kpc.  A given system will be unstable if
$\lambda/\pi r$ is less than or of order unity at any radius.  By this
criterion, all five systems are stable.}
\end{figure}

\clearpage

\begin{center}
Table 1.

Observational and cosmological parameters of the clusters.

\begin{tabular}{ccccccc}\tableline
Name  		&$T_{in}$ (keV)	&$T_{out}$ (keV)&$M_{vir}$ ($M_{\sun}$)	&$r_s$ (kpc) &$M_0$ ($M_{\sun}$)	& ref \\\tableline\tableline
Abell 1795	&2	&7.5	&$1.2\times 10^{15}$	&460	&$6.6\times 10^{14}$	&(1,2)\\
Abell 1835	&5	&11	&$1.7\times 10^{15}$	&520	&$1.1\times 10^{15}$	&(3)\\
		&4	&8	&$1.1\times 10^{15}$	&470	&$8.8\times 10^{14}$	&(4,5)\\
Abell 2052	&1.3	&3.5	&$3.9\times 10^{14}$	&340	&$2.3\times 10^{14}$	&(6)\\
Abell 2199	&1.6	&5	&$6.6\times 10^{14}$	&390	&$3.8\times 10^{14}$	&(7)\\
Abell 2390	&4	&11	&$1.7\times 10^{15}$	&520	&$1.1\times 10^{15}$	&(8)\\ 
Abell 2597	&1	&4	&$4.6\times 10^{14}$	&360	&$2.8\times 10^{14}$	&(9)\\
Hydra A		&3	&4.1	&$4.9\times 10^{14}$	&370	&$2.9\times 10^{14}$	&(10)\\
RXJ 1347.5-1145	&6	&16	&$2.5\times 10^{15}$	&600	&$1.8\times 10^{15}$	&(11)\\
Sersic 159-03	&2	&2.7	&$2.6\times 10^{14}$	&310	&$1.6\times 10^{14}$	&(12)\\
3C295		&2	&6	&$5.8\times 10^{14}$	&420	&$4.9\times 10^{14}$	&(13)\\ \tableline
\end{tabular}

References: (1) \citet{etto02}; (2) \citet{tamu01}; (3) \citet{schm01}; (4) \citet{mark02}; (5) \citet{maje02}; (6) \citet{blan01}; (7) \citet{john02}; (8) \citet{alle01b}; (9) \citet{mcna01}; (10) \citet{davi01}; (11) \citet{alle02}; (12) \citet{kaas01}; (13) \citet{alle01a}.  
\end{center}

Notes: Observed parameters ($T_{in}$ is the temperature in the innermost region of the cluster, $T_{out}$ is the temperature outside the cooling region) and inferred parameters of the NFW profile (the virial mass of the cluster, $M_{vir}$; the scale radius $r_s$ and the characteristic mass $M_0$); see \S2 on how the NFW parameters were obtained. The last column contains references to the original publications for the X-ray data on the gas temperature and electron density profiles.

\clearpage
\begin{center}
Table 2.

Best-fit parameters of the conduction model.

\begin{tabular}{ccccc}\tableline
Name  		&f (best-fit)	&$n_e$(0) (cm$^{-3}$)	&$r_c$ (kpc) &\\\tableline\tableline
Abell 1795	&0.2		&0.049		&$r_s/20$	&(see Fig. 1)\\
Abell 1835	&0.4		&0.17		&$r_s/20$	&\\
Abell 2052	&0.6		&0.089		&0		&\\
Abell 2199	&0.4		&0.074		&0		&(see Fig. 2)\\
Abell 2390	&0.3		&0.069		&$r_s/20$	&(see Fig. 3)\\ 
Abell 2597	&2.4		&0.13		&$r_s/20$	&\\
Hydra A		&1.5		&0.020		&0		&\\
RXJ 1347.5-1145	&0.3		&0.11		&$r_s/20$	&\\
Sersic 159	&5.6		&0.042		&0		&\\
3C295		&1		&0.084		&$r_s/20$	&\\ \tableline
\end{tabular}
\end{center}

Notes: $f=\kappa/\kappa_{Sp}$ is the value of the conductivity
coefficient required to reproduce the observed electron density,
$n_e$(0) is the corresponding model electron density in the center,
and $r_c$ is the core radius adjusted to fit the observed temperature
profile, for which we only consider two values (see \S3).

\end{document}